\documentclass[twocolumn]{jpsj2} 
\title{Why Does Zipf's Law Break Down in 
Rank-Size Distribution of Cities?}

\author{Hiroto \textsc{KUNINAKA}
\thanks{E-mail address: kuninaka@phys.chuo-u.ac.jp}
and Mitsugu \textsc{Matsushita}}

\inst{Department of Physics, Chuo University, Kasuga, Bunkyo-ku, Tokyo 112-8551}

\abst{We study rank-size distribution of cities in Japan 
on the basis of data analysis and computer simulation. 
From the census data after World War II, 
we find that the rank-size distribution of cities 
is composed of two parts, each of which has independent 
power exponent. In addition, the power exponent of the head part 
of the distribution changes in time and Zipf's law holds 
only in a restricted period. We show that Zipf's law broke down 
due to both of Showa and Heisei great mergers and recovered 
due to population growth in middle-sized cities after the great 
Showa merger. 
}

\kword{population, rank-size distribution, 
power-law distribution, lognormal distribution, Zipf's law, 
Gibrat's law, agent-based model}

\begin{document}
\maketitle
\section{Introduction}
Many empirical data which obey power-law distribution 
can often be observed in both natural and social sciences.\cite{newman,aaron} 
For example, the size distribution of lunar craters\cite{lunar}, 
the relation between frequency and magnitude of earthquakes 
\cite{G-R}, the size distribution of islands\cite{island} 
and the cumulative probability distribution of personal income in Japan
\cite{aoyama} are known to obey power-law distribution. 

On the other hand, many researches have reported that 
empirical data obeying lognormal distribution are abundant around us. 
The lognormal distribution has the form  
\begin{equation}\label{lognormal}
N(z)=\frac{1}{\sqrt{2\pi\sigma^2}z} 
\exp \left( -\frac{[\ln(z/T)]^2}{2\sigma^2} \right),
\end{equation}
where $\sigma$ and $T$ are dispersion and average, respectively. 
For example, the fragmentation of glass rods\cite{glass},
income distribution of families and single individuals in U.S.\cite{inc},
the size distribution of barchan dunes\cite{barchan}, 
food fragmentation by chewing\cite{chew} 
and the duration of disability for aged people\cite{mori} 
are believed to obey Eq.(\ref{lognormal}) approximately.

The origin of those characteristic distributions is explained 
by the random multiplicative process which is often used 
to mimic the growth process of living organisms\cite{ln}. 
Let $X_{i}$ be a physical quantity at time step $i$ and 
suppose that its growth process is governed 
by the following relation: $X_{i}=\alpha_{i-1}X_{i-1}$. Here, 
$\alpha_{i}$ is the growth rate at the time step $i$. 
At the $m$-th step, $X_{m}$ can be written as 
$X_{m} = X_{0} \prod_{i=0}^{m-1} \alpha_{i}$, 
where $X_{0}$ is the initial quantity. 
Thus, when we assume that $\alpha_{i}$ is a random variable 
independent of $X_{i}$ and $m$ is sufficiently large, 
$\log X_{i}$ obeys the normal distribution due to the central limit
theorem, which entails the lognormal distribution of $X_{i}$. 
This process is often called Gibrat's process or Gibrat's law, which 
is so common to many complex systems that one may say that the default
distribution is the lognormal distribution. 
Indeed, if we introduce additional term, such as a noise term, 
into the Gibrat's process, 
$X_{i}$ obeys power-law distributions. \cite{takayasu,tomita}


This paper focuses on the population distribution of cities in Japan. 
The population distribution within a given region sometimes 
shows power-law behavior. 
Auerbach firstly reported that the rank-size distribution for 
population of cities obeys a power-law distribution.\cite{auerbach} 
This means that when we order the cities by population and plot the 
rank $R(x)$ against its corresponding population $x$,  
the relation between $R(x)$ and $x$ can be approximated 
by
\begin{equation}\label{eq1}
\log R(x) = a - b \log x, 
\end{equation}
where $a$ and $b$ are fitting parameters. 
As for other municipalities, for example, Sasaki {\it et al.} recently 
reported that the rank-size distributions of towns and villages 
can be well approximated by lognormal distributions while that of cities 
obeys power-law distribution in Japan\cite{sasaki}.

Zipf reported that the power exponent $b$ is approximately $1$ 
in the case of cities, so that the special case $b=1$ is generally 
called  Zipf's law.\cite{zipf} Since 
many empirical data, such as the frequency of words in a literature and  
the income of companies\cite{zipf2}, obey Zipf's law, 
it is believed to be universal regularity.

\begin{figure}[htb]
\begin{center}
\includegraphics[width=0.4\textwidth]{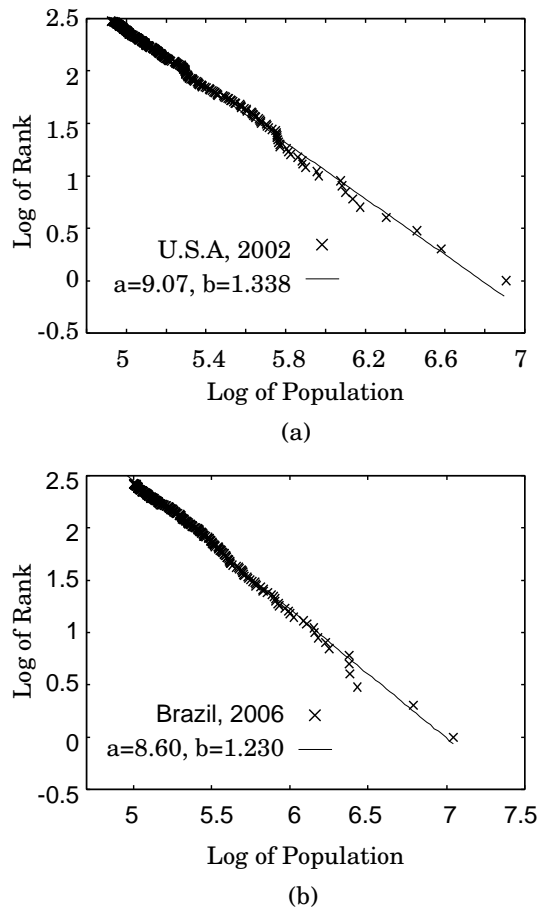}
\end{center}
\caption{Rank-size distributions for population of cities 
(a) in U.S.A, 2002 and (b) in Brazil, 2006.}
\label{fig1}
\end{figure}

However, we can easily find that Zipf's law in population 
distribution of cities is not universal.\cite{kwok} 
For example, Figs. \ref{fig1} (a) and (b) show the rank-size 
distribution of top 300 cities of 
U.S.A. in 2002 and that of top 267 cities of Brazil in 2006, respectively. 
Power exponents are $b=1.338 \pm 0.004$ and $b=1.230 \pm 0.005$, 
respectively, which are different from $b=1$ predicted by Zipf's law, 
although the distributions obey power-law behavior. 
Here, we obtain those power exponents by the least square linear 
regression after plotting log of rank against log of population. 
In addition, we often find that the rank-size distribution 
does not exhibit power-law behavior.\cite{kwok}  
Even if the rank-size distribution obeys power-law distribution, 
it is easily expected that power exponent $b$ changes 
in time due to migration, a change of birth rate, and so on. 


Some stochastic models have been proposed to explain Zipf's law.
For example, Simon's model explains the emergence of Zipf's law 
in the rank-frequency distribution of words in literature.~\cite{simon} 
The point of this model is that, for adding a new word to a text, 
the probability that a word is repeated is proportional to 
the number of its previous occurrence. 
This model explains that the rank-frequency distribution obeys 
power-law distribution with the exponent less than or equal to unity, 
which depends on the probability to choose the newly added word. 
On the other hand, Cancho has developed a model to explain the case 
that the exponent becomes larger than unity.~\cite{cancho} 
Recently, by modifying Simon's model, 
Zanette and Montemurro have developed a more realistic model 
to reproduce Zipf's law in the rank-frequency distribution of words, 
which explains the exponents in some cases of different languages. 
~\cite{zanette}.

In this paper, we investigate the time evolution of 
the rank-size distribution of cities in Japan 
to show how the power exponent $b$ changes after World War II. 
In addition, we show that Zipf's law holds only in a restricted period 
to explain why Zipf's law breaks down in Japan from 
our results of data analysis and simulation. 
Our data analysis is based on the census data 
from 1950 to 2006 which obtained from Statistics Bureau, 
Ministry of International Affairs and Communications, Japan\cite{toukei}, 
and data book from Japan Statistical Association.\cite{soq} 

The organization of this paper is as follows. 
In the next section, we show our data analyses about the time evolution 
of the rank-size distribution for population of cities and the 
power exponent of its head part. 
Section 3 is devoted to modelling of population migration to explain 
the time evolution of the power exponent. 
In \S 4, we discuss our results of data analyses and simulation. 
The final section summarizes our results. 

\section{Data Analyses}

\begin{figure}[htb]
\begin{center}
\includegraphics[width=0.4\textwidth]{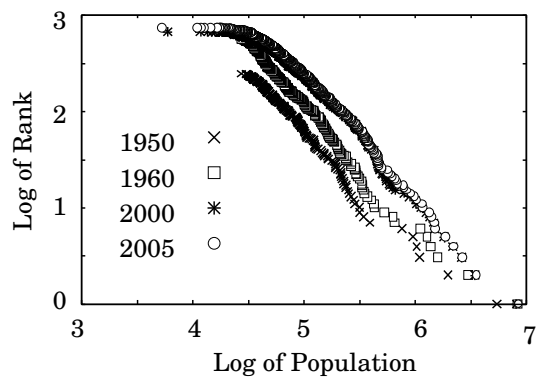}
\end{center}
\caption{Rank-size distribution of cities in Japan from 1950 to 2005.}
\label{fig2}
\end{figure}

Figure \ref{fig2} shows the rank-size distributions for cities 
of Japan in 1950, 1960, 2000 and 2005, respectively. 
In each year, the rank-size distribution can be divided 
into two parts. For example, the distributions in 2000 and 2005 are 
clearly divided into two parts around $5.0 \times 10^{5}$ in population. 
Thus, the head and the tail part of each distribution 
can be fitted by discrete power-law distribution functions. 

The slopes of head parts of the rank-size distributions 
change significantly from 1950 to 1960. 
This is mainly due to the fact that the number of cities drastically 
increased from 248 to 565 in the the great Showa merger from 1955 to 1960. 
From 2000 to 2005, the slope of head parts slightly increases 
from $1.027 \pm 0.004$  to $1.080 \pm 0.004$, 
although the two distributions globally seem to be similar. 
Also in this case, 
the change of slopes of head parts is affected by the increase of 
the number of cities due to the great Heisei merger from 2000. 
Thus, the power exponent of the distribution changes easily by 
the great merger of municipalities. 

\begin{figure}[htb]
\begin{center}
\includegraphics[width=0.4\textwidth]{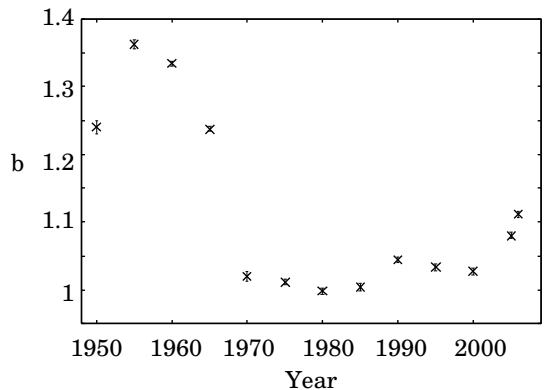}
\end{center}
\caption{Time evolution of power exponent $b$ from 1950 to 2006.}
\label{fig3}
\end{figure}

Next, we investigate how the power exponent of head part of 
the rank-size distribution changed in time after World War II. 
Figure \ref{fig3} shows the time evolution of the power exponent $b$ 
from 1950 to 2006. Error bars which are almost invisible on data marks 
are standard deviation obtained by the least-squares linear regression.
This figure shows that the power exponent $b$ 
drastically changes during the two great mergers both in Showa 
and Heisei era. 
After the great Showa merger finished in 1960, 
the power exponent $b$ shows monotonic decrease and approaches unity. 
The power exponent $b$ keeps the value near unity until the great 
Heisei merger starts in 2000. Thus, it is shown that Zipf's law holds 
only in the period from 1970 to 2000 in Japan. 

Here, we should comment on the fitting range to obtain $b$. 
As we can see in Fig.\ref{fig2}, the range of the head part is 
not so large at each year. Thus, all the values in Fig.\ref{fig3} 
are obtained by regression within about one order of magnitude.

To explain the relaxation of $b$ to unity, we investigate the time 
evolution for the growth rate of population. 
To calculate the growth rate, at first, we categorize cities 
into some groups. 
The $n$-th group ($2 \le n \le 4$) is composed of $80$ cities, 
the rank of which ranges from $80n-139$ to $80n-60$ at a given year, 
while the first group ($n=1$) is composed of $20$ cities, 
the rank of which ranges from $1$ to $20$. 
Note that the constituents of each group changes because the rank 
of cities usually changes at every census year. 

\begin{figure}[htb]
\begin{center}
\includegraphics[width=0.4\textwidth]{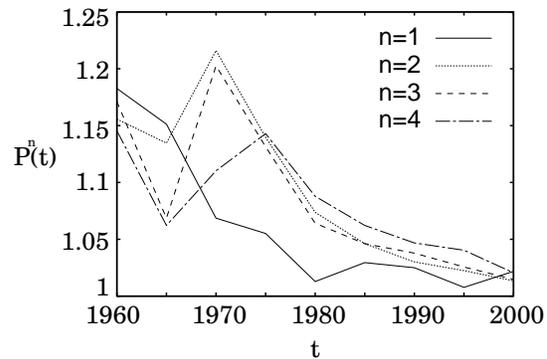}
\end{center}
\caption{Time evolution of growth rate from 1960 to 2000.}
\label{fig4}
\end{figure}

We define the growth rate $P^{n}(t)$ of the $n$-th group 
at a census year $t$ as 
\begin{eqnarray}\label{eq2}
P^{n}(t)=\frac{\sum_{i=1}^{n_{c}}x^{n}_{i}(t)}
{\sum_{i=1}^{n_{c}}x^{n}_{i}(t-\Delta t)},\\
n_{c}=
\begin{cases}
20 & (n=1)\\
80 & (n=2,3,4),
\end{cases}\notag
\end{eqnarray}
where $x_{i}^{n}(t)$ is the population of the $i$-th city which belongs to 
the $n$-th group, and $\Delta t$ is taken as $\Delta t=5$ 
which is the interval between two successive census years in Japan. 
Figure \ref{fig4} is the time evolution of the growth rate $P^{n}(t)$ of 
each group from 1960 to 2000. The growth rate of the first group 
shows global decrease, while those of other groups have apparent 
peaks in 1970 or 1975. This may be attributed to the following two 
factors: (i) the migration from the big cities to 
their satellite cities or the countryside, 
such as ``U turn phenomena'' or ``I turn phenomena'', 
which is remarkable after 1970\cite{migration} and  
(ii) population increase due to the second baby boom in the first half 
of the 1970s. 

From these results, we can understand the time evolution of 
the power exponent $b$ in Fig. \ref{fig3} by the following scenario: 
\begin{enumerate}
\item Before the great Showa merger starts, the power exponent $b$ has 
the value near unity. 
\item Due to the increase of the number of cities by the great 
Showa merger, the power exponent $b$ increases. 
\item After the merger, under the circumstance 
that the increase of the number of cities is not so large, 
the population of cities whose ranks range 
from 20 to 260 increases, which results in the decrease of $b$.  
\item The power exponent $b$ remains the value near unity until the 
great Heisei merger starts in 2000. 
\end{enumerate}

Here we would like to comment on why the period in which Zipf's law 
held continued for about 30 years. 
The head part of the rank-size distribution of cities consists of 
the groups with $n \ge 2$. 
From Fig.~\ref{fig4}, we can easily find that the growth rates 
of these groups have almost the same value after 1975. 
In addition, the number of cities showed slow increase after 1960, 
while it had shown fast increase between 1950 and 1960~\cite{toukei} 
(see Table.~\ref{noc}). 
This may cause the stability of the power exponent after Zipf's law
holds and prevent the exponent from taking the value less than unity. 

\begin{table}[tb]
\caption{Number of cities of each year.}
\label{noc}
\begin{tabular}{cc}
\hline
Year & Number of cities \\
\hline
1950 & 254 \\
1960 & 561 \\
1970 & 588 \\
1980 & 647 \\
1990 & 656 \\
\hline
\end{tabular}
\end{table}

\section{Modelling on Population Migration}\label{mpm}
In this section, we construct a model for the population migration 
to reproduce the increase of $b$ due to the merger of 
municipalities and its convergence to unity after the merger. 
Our model is based on an agent-based model 
which consists of $3500$ sites corresponding to all the municipalities. 
Each site has a uniform random number between $0$ and $1$ 
as the initial population. 
The basic procedure of one simulation step is summarized as follows: 
\begin{enumerate}
    \item We randomly choose a source site $m$ with the population $N_{m}$. 
    \item We choose a group of sites, $G_{N<N_{m}}$ or $G_{N>N_{m}}$,   
which are the groups of the sites whose population $N$ 
are less and more than $N_{m}$, respectively. 
The probability to choose $G_{N<N_{m}}$ is $\alpha$ (migration parameter) 
while that to choose $G_{N>N_{m}}$ is $1-\alpha$. 

    \item Among the group of sites chosen in the previous step, 
we randomly choose the destination site $n$ for migration. 
    \item $P_{mn}$ percent of $N_{m}$ are transferred to the site $n$,
	  so that the populations of sites $m$ and $n$ vary in quantity 
as $N_{m}-P_{mn}N_{m}$ and $N_{n}+P_{mn}N_{m}$, respectively. 
\end{enumerate} 
In the second step, the migration parameter $\alpha$ is introduced 
to describe the tendency that people migrate to less populated area 
from large cities which was evident after the high economic growth 
from 1960 to early 1970s.\cite{migration} 
In addition, $P_{mn}$ is randomly chosen 
in the range from $0$ to $20$. 
We iterate this procedure $10^{6}$ times in our simulation. 
Sample average is taken over 10 different initial population 
distributions for all the sites. 

When the population of a given site becomes larger than $0.95$, 
we regard the site as a city. Once a site is promoted to a city, 
the site will not be demoted to a smaller municipality 
such as towns and villages. This rule corresponds 
to a part of the Local Autonomy Law of Japan  
which says that municipalities must have a population of 
$50,000$ or more to be promoted to cities\cite{law}. 
Our model does not distinguish between towns and cities. 
Thus, if a site does not belong to cities, 
we henceforth call the site as a ``town''. 

After the first migration of $10^{6}$ simulation steps, 
we merge some municipalities according to the following procedure. 
At first, we randomly choose two sites to merge among all the sites. 
When both of them are not cities, we merge them to produce a new city 
if the sum of those populations becomes larger than $0.95$, 
while we merge them to produce a town if the sum is less than $0.95$. 
On the other hand, when at least one site is a city, we merge those two 
sites with the probability $\beta=0.5$ to become a new city. 
The probability $\beta$ is introduced due to the fact that the frequency 
of the merger of towns was much larger than that of cities.  
We iterate this merging process  until the number of cities increases 
by $77$ on average rather than that 
when the first migration stage is finished. 
In our model, the increase of the number of cities affects 
the power exponent after the merger. In general, 
the power exponent increases with the increase of the number of cities 
generated by the merger. 

\section{Simulation Results}
\begin{figure}[t]
\begin{center}
\includegraphics[width=0.4\textwidth]{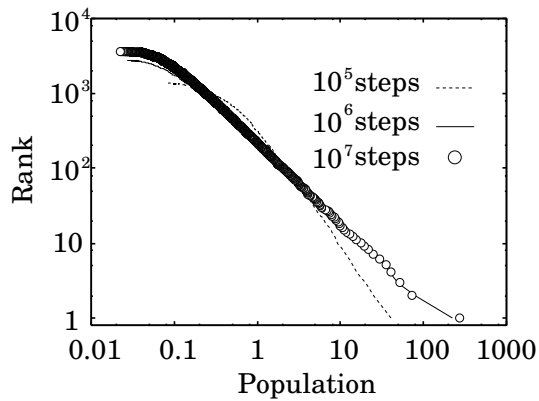}
\end{center}
\caption{Time evolution of rank-size distribution of cities without 
 merging process.}
\label{conv}
\end{figure}

At first, we investigate the convergence of 
the rank-size distribution of cities generated by our model. 
Figure \ref{conv} shows the rank-size distributions of cities 
at $10^{5}$, $10^{6}$, and $10^{7}$ simulation steps, respectively. 
To obtain these results, the value of $\alpha$ is fixed at $\alpha=0.3$. 
This figure shows that the rank-size distribution converges 
to the stationary power-law distribution 
with the power exponent $b=1.012 \pm 0.002$. 
When the number of sites is more than $3500$, 
our model needs longer simulation steps for the convergence 
to the power-law distribution with $b=1$. 
Thus, our model can reproduce the power-law distribution of 
cities which converges to Zipf's law. 
However, in our model, the number of cities keeps increasing 
after the power exponent becomes $b=1$, 
which slightly increases the power exponent.

\begin{figure}[t]
\begin{center}
\includegraphics[width=0.4\textwidth]{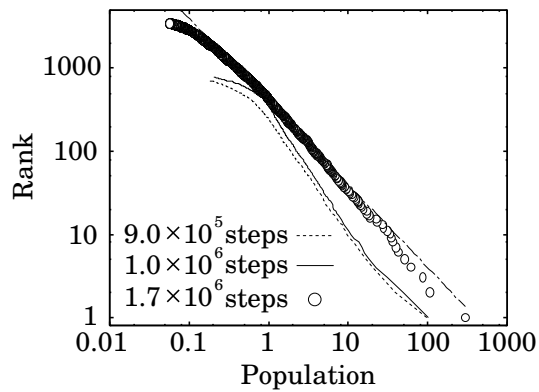}
\end{center}
\caption{Time evolution of rank-size distribution of cities with 
 merging process.}
\label{jikan}
\end{figure}

Secondly, we investigate how the great merger affects 
the rank-size distributions of cities through 
the time evolution of the power exponent $b$. 
In this simulation, we carry out the first population migration 
of $10^6$ simulation steps. After that, we merge some of those sites, 
followed by the second population migration of $7 \times 10^5$ 
simulation steps. 
Figure \ref{jikan} shows the time evolution of the rank-size
distribution of cities. 
The dotted line shows the distribution after the first migration 
stage was finished. The solid line shows the distribution after 
a merger of 200 sites. The open circles show the distribution 
after the second migration stage was finished, which can be fitted 
by the power-law distribution with the exponent $b=1.081 \pm 0.001$ 
denoted by the dash-dotted line. Here we find that the distribution 
approaches the power-law 
distribution with the exponent $b=1$ after the merger.

\begin{figure}[t]
\begin{center}
\includegraphics[width=0.4\textwidth]{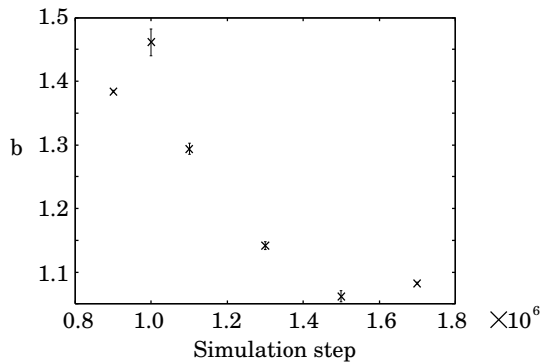}
\end{center}
\caption{Time evolution of power exponent $b$.}
\label{bt}
\end{figure}

We show the relation between the power exponent $b$ and 
the simulation step in Fig.~\ref{bt}. 
Error bars which are almost invisible on a few data marks 
are standard deviation obtained by the least-squares linear regression. 
Data point at $10^{6}$ steps shows the power exponent $b$ after the  
merger has finished. 
We find that $b$ converges to unity after the increase of $b$ 
due to the merger. Thus, our model can reproduce 
the time evolution of $b$ qualitatively.


\begin{figure}[t]
\begin{center}
\includegraphics[width=0.4\textwidth]{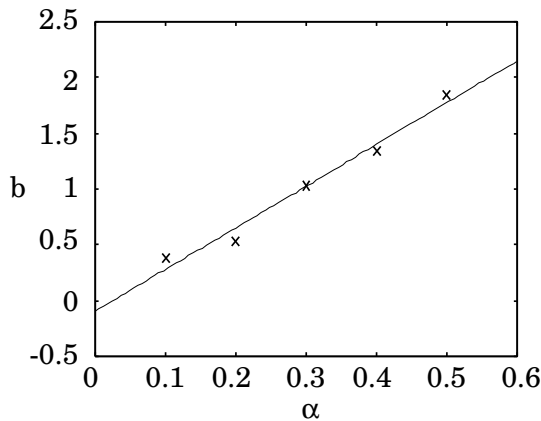}
\end{center}
\caption{Relation between $\alpha$ and $b$.}
\label{b_vs_a}
\end{figure}

Finally, we investigate how $\alpha$ affects the final distribution. 
Figure \ref{b_vs_a} shows the relation between $\alpha$ and 
the power exponent $b$ at $10^{6}$ simulation steps. 
The solid line is  the regression line: $b=3.7 \alpha - 0.09$.
This result indicates that $\alpha$ determines the power exponent $b$ 
of the final power-law distribution. For the convergence to Zipf's law, 
this model requires $\alpha=0.3$. 

Here we would like to comment on the effect of 
initial population distribution on the final distribution. 
When we give $N_{i}=1.0$ for initial value of all the sites, 
the power exponents of the resulting distributions do not show 
large difference. 

%
%
\section{Discussion}
Let us discuss our results. 
From Fig.\ref{fig3}, we find that Zipf's law holds for 25 years and 
breaks down due to the great Heisei merger. 
Naturally arises a question whether Zipf's law held 
also before 1950. 
However, during World War II, the number and distribution of people 
must have shown large fluctuation due to the great air campaigns 
against large cities such as Tokyo and Osaka, 
and evacuations from large cities to countrysides. 
Under a circumstance that the population distribution is unstable, 
it may be of little importance in discussing whether Zipf's law holds 
because there is a possibility that the distribution no more obeys 
power-law one. 

\begin{figure}[htb]
\begin{center}
\includegraphics[width=0.4\textwidth]{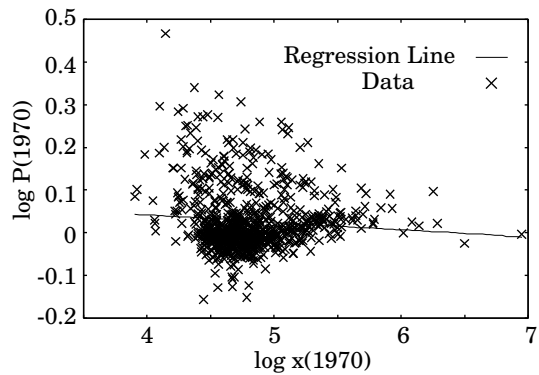}
\end{center}
\caption{Relation between log of growth rate and log of population 
for cities from 1965 to 1970. Solid line is Eq.(\ref{regl})}
\label{fig9}
\end{figure}

We have found that the power exponent $b$ approached unity after 
the great Showa merger had finished. 
Some theoretical explanations for the emergence of Zipf's law have 
been proposed in literature\cite{gabaix,kawamura,simon}. 
Among them, 
Gabaix showed that 
Gibrat's law in the population growth of each city is necessary 
for the emergence of Zipf's law.\cite {gabaix,ionnides} 
Here, Gibrat's law means that different cities grow randomly 
with the growth rate independent of the population of cities. 
We investigated the relation between the growth rate 
$P(t) \equiv x(t)/x(t-\Delta t)$ 
and the population $x(t)$ for all cities at $t=1970$ (Fig.\ref{fig9}). 
The reason why we focus on $t=1970$ is that the convergence 
to Zipf's law can be seen in this period (Fig.\ref{fig3}).
Here, we can run the regression,
\begin{equation}\label{regl}
\log P(1970) = 0.11-(0.017 \pm 0.007) \log x(1970),
\end{equation}
which has a slight slope, although it is supposed to 
become $0$ if Gibrat's law holds. 
In addition, the dispersion of growth rate becomes rather large 
around $\log x(1970) = 4.5$, so that we cannot clearly see 
whether Gibrat's law holds or not.  
In the case of all municipalities, Sasaki {\it et al.} reported 
that the slope becomes almost $0$ from 2000 to 2005, although 
it has a slight slope.\cite{sasaki} 

As we referred in \S 1, a random multiplicative process with 
Gibrat's law generates lognormal distribution. 
The rank-size distribution for population in all municipalities 
shows the double-Pareto distribution, which consists of lognormal 
body with power-law tail\cite{mitz,sasaki,tomita}. 
Thus, it is no wonder that the regression line for 
the relation between the growth rate and the population has 
a non-zero slope. 
Because the rank-size distribution of cities is the tail part of 
that of municipalities, it may have a non-zero slope. 
Thus, in the case of population of cities, Gibrat's law may be 
just necessary condition for the emergence of Zipf's law. 

To obtain the power exponent $b$ of rank-size distributions 
for cities, 
we adopt the least-squares linear regression to fit those distribution 
functions by Eq.(\ref{eq1}). 
Although this method is used frequently in literature, it is known that
the method has several problems.\cite{aaron} 
To obtain more reliable estimate for $b$, other estimation methods 
such as the maximum likelihood method\cite{aaron} may be better. 

In \S 3, we have constructed the agent-based model to 
explain the emergence and the breakdown of Zipf's law 
in the rank-size distribution of cities. 
This model can reproduce Zipf's law which is observed 
in the process that many entities exchange physical quantities among them. 
We often find Zipf's law in some phenomena without 
an apparent exchange process such as word frequencies in literature 
and the relation between the frequency and the magnitude of earthquakes. 
However, even for these cases there may be some hidden exchange 
processes such as words in/out of fashion and 
the accumulation/relaxation of the crust stress due to the plate tectonic 
movement. Moreover, exchange processes are almost universal in the
economic world as well as social world. Hence we believe that the
present model can be applicable to other problems as well. 

In \S 4, we carried out a simulation of population 
migration to explain the time evolution of the power exponent $b$ 
in Fig.~\ref{fig3}. 
In Fig.~\ref{jikan}, after the merger of municipalities, 
the rank-size distribution shifts towards upper direction 
in all the region.
If we use the value of $\beta$ smaller than $0.5$, 
the distribution shifts towards upper direction 
in the region whose population is less than about $1.8$. 
Consequently, small value of $\beta$ causes a decrease of the range in which 
the distribution can be fitted by a single power-law distribution. 

The rank-size distribution of all municipalities has 
a lognormal body and a power-law tail\cite{sasaki}, 
which is observed also in our simulation. 
This type of distribution can be observed in the agent-based simulation 
of exchanging quantities on the small-world network\cite{souma}, 
which implies the possibility that the population migration network 
may have a small-world structure. To clarify the relevance, 
we need to analyse the population migration network 
between municipalities in detail.

%
%
\section{Concluding Remarks}
In conclusion, we have investigated the time evolution of 
the rank-size distribution for population of cities to show 
how the power exponent changes in time. 
The rank-size distribution shows that power-law behavior and the 
time evolution of the power exponent drastically changes 
when the great merger of municipalities occurs. 
After the great Showa merger finished, the power exponent 
converged to unity, which means that Zipf's law holds. 
We have explained the change of the power exponent by 
the growth rates of the categorized groups of cities in the 
point of view of migration.


%
%
We would like to thank M. Katori, T. Nakano, S. Tomita, N. Kobayashi, 
Y. Sasaki and T. Miyazaki for useful discussions. 
We would also like to thank Y. Aruka, A. Namatame and H. Hayakawa 
for their useful comments. 
Numerical computation was partially carried out at the Yukawa Institute 
Computer Facility. 
A part of this work is supported by a Grant-in-Aid for Young Scientists 
(B) of the Ministry of Education, 
Culture, Sports, Science and Technology (MEXT), Japan (Grant No.20740226). 

\end{document}